\documentclass[journal=ancac3,manuscript=article,layout=twocolumn]{achemso}

\usepackage[version=3]{mhchem} 

\author{Entao Yang}
\affiliation{Department of Chemical \& Biomolecular Engineering, University of Pennsylvania, Philadelphia, PA 19104, USA}
\author{James F. Pressly} 
\affiliation{Department of Materials Science \& Engineering, University of Pennsylvania, Philadelphia, PA 19104, USA}
\author{Bharath Natarajan}
\affiliation{ExxonMobil Technology and Engineering Company, Annandale, NJ 08801, USA}
\author{Robert Colby}
\affiliation{ExxonMobil Technology and Engineering Company, Annandale, NJ 08801, USA}
\author{Karen I. Winey} 
\email{winey@seas.upenn.edu}
\affiliation{Department of Materials Science \& Engineering, University of Pennsylvania, Philadelphia, PA 19104, USA}
\alsoaffiliation{Department of Chemical \& Biomolecular Engineering, University of Pennsylvania, Philadelphia, PA 19104, USA}
\author{Robert A. Riggleman} 
\email{rrig@seas.upenn.edu}
\affiliation{Department of Chemical \& Biomolecular Engineering, University of Pennsylvania, Philadelphia, PA 19104, USA}

\title[An \textsf{ancac3} demo]
{Understanding Creep Suppression Mechanism in Polymer Nanocomposites through Machine Learning}


\begin{document}


\begin{abstract}
While recent efforts have shown how local structure plays an essential role in the dynamic heterogeneity of homogeneous glass-forming materials, systems containing interfaces such as thin films or composite materials remain poorly understood. 
It is known that interfaces perturb the molecular packing nearby, however, numerous studies show the dynamics are modified over a much larger range. 
Here, we examine the dynamics in polymer nanocomposites (PNCs) using a combination of simulations and experiments and quantitatively separate the role of polymer packing from other effects on the dynamics, as a function of distance from the nanoparticle surfaces. 
After showing good qualitative agreement between the simulations and experiments in glassy structure and creep compliance, we use a recently developed machine learning technique to decompose polymer dynamics in our simulated PNCs into structure-dependent and structure-independent processes.
With this decomposition, the free energy barrier for polymer rearrangement can be described as a combination of packing-dependent and packing-independent barriers.
We find both barriers are higher near nanoparticles and decrease with applied stress, quantitatively demonstrating that the slow interfacial dynamics is not solely due to polymer packing differences, but also the change of structure-dynamics relationships. 
Finally, we present how this decomposition can be used to accurately predict strain-time creep curves for PNCs from their static configuration, providing additional insights into the effects of polymer-nanoparticle interfaces on creep suppression in PNCs.
\end{abstract}

\section{Introduction}

Due to the continuous growth of the world population, the global construction market is predicted to increase $85\%$ by 2030 compared to 2015\cite{GlobalConstructionPerspectives2015}.
Flourishing construction will strain materials supply streams, including wood, sand used for cement, and iron ore for steel, challenging the sustainable utilization of these limited natural resources.
To fill the gap in materials supply and demand within the infrastructure sector, development of new materials for structural applications is desired.
Polymer composites are a promising candidate due to their light weight and corrosion resistance.\cite{Das2001,Shrivastava2016}
However, polymers, particularly recycling-friendly thermoplastics, tend to creep under long-term external loads,\cite{Raghavan1997} limiting their application as infrastructural materials, which typically require a service life of 50 to 100 years.\cite{Gulvanessian2002a}

Studies have shown that adding nanoparticles (NPs) with neutral or attractive polymer-NP interactions can significantly change the mechanical properties of polymer matrices, including increasing tensile strength and average shear and Young's moduli, altering polymers' nonaffine displacement field during deformation, and rendering the material less fragile.\cite{knauert2007effect,Papakonstantopoulos2005,Papakonstantopoulos2007,Kumar2017,Buitrago2020}
While some studies show that polymer nanocomposites (PNCs) have a better resistance to creep,\cite{Zhang2004a,Ranade2005,Yang2006,Riggleman2009,Buitrago2020} the mechanism of creep suppression, including the role of NP size, loading, and polymer-NP interactions, remains unclear.

The presence of an interfacial layer where polymer monomers exhibit decreased mobility is known to be critical to mechanical reinforcement in PNCs.\cite{Papakonstantopoulos2005,Cheng2016,Baeza2016}
In both experiments\cite{Jouault2013,Holt2014,Cheng2017} and simulations,\cite{Starr2002,Zhang2010,Starr2016,Emamy2018} attractive NPs have been shown to create a layer of slowed polymer dynamics, often several orders of magnitude slower than in bulk polymers.
Simulation studies have shown that these local dynamical changes are not solely attributable to denser packing near the NP surface, as evidenced by the thickness of the structurally affected region decreasing with cooling while the thickness of the dynamically affected region increases.\cite{Zhang2019}
Recent work also suggests that the presence of NPs can slow down the polymer diffusion, which persists far beyond the length scale where the polymer conformation are modified.\cite{bailey2020polymer}
Mirigian et al. have formulated a force-level theory, which divides the free energy of relaxation into a local barrier based on Nonlinear Langevin Equation (NLE) theory\cite{Schweizer2005} and a long-range barrier described by an elastic continuum, accurately describing relaxation progress in both bulk supercooled liquids and free-standing films.\cite{Mirigian2013,Mirigian2015}
These results suggest that changes in the segmental packing are not the only factor controlling the dynamical gradient around NPs.
Thus, investigating the relationship between polymer dynamics and structure and how the relationship changes as a function of different conditions (such as applied stress, NP size, and polymer-NP interaction) is important to understanding not only creep suppression, but also the underlying mechanisms behind NP reinforcement in PNCs.

Over the past several years, an application of machine learning has been proposed in disordered materials to directly connect a particle's (a polymer monomer in this work) local-structure features, termed as softness, to its probability of rearrangement (defined as having a relatively large non-affine local displacement. See Methods for more details).\cite{Cubuk2015,Schoenholz2016,Schoenholz2017}
This new method easily measures the structure in amorphous materials and estimates its effect on monomer-level dynamics.
Softness expanded our understanding of glassy materials, including the universal yield strain\cite{Cubuk2017}, the aging process\cite{Schoenholz2017}, and the structural initiation of shear banding.\cite{Ivancic2019,Yang2020}
Using softness, Liu and coworkers decomposed dynamics in both bulk glasses and glassy polymer thin films\cite{Schoenholz2016,Sussman2017} into structure-dependent and structure-independent components and found that the slowing of dynamics near the glass transition in bulk glasses is associated with structural changes, whereas enhanced dynamics near free surfaces in glassy thin films are dominated by a structure-independent mechanism.
While most of these softness analyses are applied to homogeneous glassy systems, our recent work has shown that softness can also be used in quantifying the structural features (i.e., local packing) of polymers in the interfacial region near NPs,\cite{Yang2020} enabling us to develop a dynamical decomposition model for polymers in PNCs.

In this work, we simulate neat polymers and dispersed nanoparticle PNCs with attractive polymer-NP interactions (strong interaction) over a range of NP loadings and two NP sizes.
The creep suppression in these systems was compared to experimental P2VP-silica composites to verify the simulations accurately describe real-world systems.
We also vary polymer-NP interactions in the model PNCs with a $10\ vol\%$ loading of NPs to study the effect of interaction strength on creep suppression.
After establishing good agreement between the simulated and experimental nanocomposite systems, we demonstrate how polymer dynamics in PNCs can be decomposed into structure-dependent and structure-independent processes and how this relation holds within the constant strain rate regime.
We refer to this relation as the dynamical decomposition model in this work.  
With this decomposition, we prove that besides the modified local packing, the relation between structure and dynamics are also changed near NPs, leading to the slow interfacial dynamics. 
Finally, we show a potential application of this decomposition model in predicting PNCs' strain response directly from the structure information of an undeformed sample.

\section{Results and Discussion}
\begin{figure}[tbhp]
\centering
\includegraphics[width=1\linewidth]{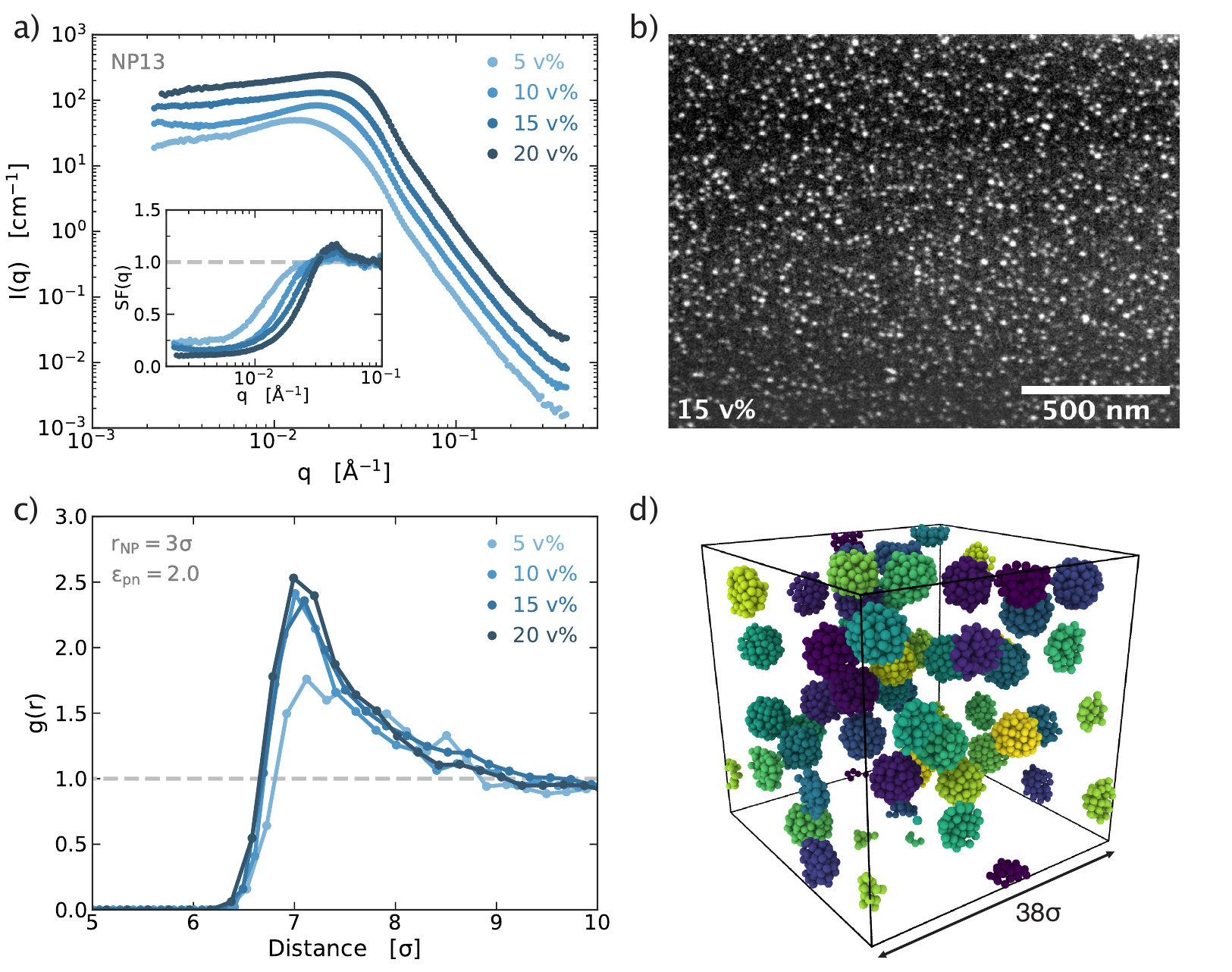}
\caption{\textbf{a)} SAXS measurements of NP13 composites showing excellent NP dispersion with an inset showing the NP structure factor.
\textbf{b)} Representative SEM image of the $15\ vol\%$ NP13 PNCs.
\textbf{c)} Pair distribution functions of NPs in simulated PNCs with small attractive NPs ($r = 3 \sigma$) showing good dispersion.
\textbf{d)} Visualization of the NP distribution in the simulated PNCs with small attractive NPs ($10\ vol\%$ NP). NPs are colored to ease differentiation.}
\label{fig:np_disp_exp}
\end{figure}

\textbf{Nanoparticle Dispersion} 
We begin our study from measuring NP dispersion state in both experimental and simulated PNCs.
Figures \ref{fig:np_disp_exp}a and \ref{fig:np_disp_exp}b contain small angle X-ray scattering (SAXS) measurements and SEM images demonstrating excellent dispersion of the $13\ nm$ diameter silica nanoparticles (NP13) within the P2VP matrix (see Figure S1 for $52\ nm$ diameter, NP52, composites).
In both the NP13 and NP52 composites, scattering in the high-q regime ($q >0.03$ \AA$^{-1}$ and $0.007$ \AA$^{-1}$, for NP13 and NP52 composites, respectively) is identical for all silica loadings.
Likewise, in the low-q regime, the scattering intensity plateaus in all systems.
This behavior is characteristic of well-dispersed NPs.
The structure factor scattering shown in the insets to Figures  \ref{fig:np_disp_exp}a and S1a further emphasize the well-dispersed nature of the composite systems, with the structure factor intensity settling at 1 with no large or sharp peaks. 
The SEM images in Figure \ref{fig:np_disp_exp}b and S1b visually confirm NP dispersion for the $15\ vol\%$ NP13 and NP52 systems, respectively.

The well-dispersed NPs observed in the experimental composite systems agree qualitatively with the nanoparticle dispersion observed in the strongly attractive ($\epsilon_{\mathit{pn}} = 2.0$) polymer-NP interaction simulations.
Figure \ref{fig:np_disp_exp}c shows the pair distribution functions, $g(r)$, of the small NPs within the composites while Figure \ref{fig:np_disp_exp}d shows a visualization of the nanoparticle dispersion within the simulation box.
(See Figures S1c and S1d for large NP composites.)
The single, relatively weak peak in $g(r)$ indicates weak ordering of the nanoparticles, suggesting good overall dispersion.
Additionally, we observed well-dispersed NPs in composite simulations with neutral interactions ($\epsilon_{\mathit{pn}} = 1.0$) and significant aggregation in simulations with weak interactions ($\epsilon_{\mathit{pn}} = 0.5$), as expected (see Figures S1e--S1h).

\begin{figure*}[tb]
    \centering
    \includegraphics[width=\linewidth]{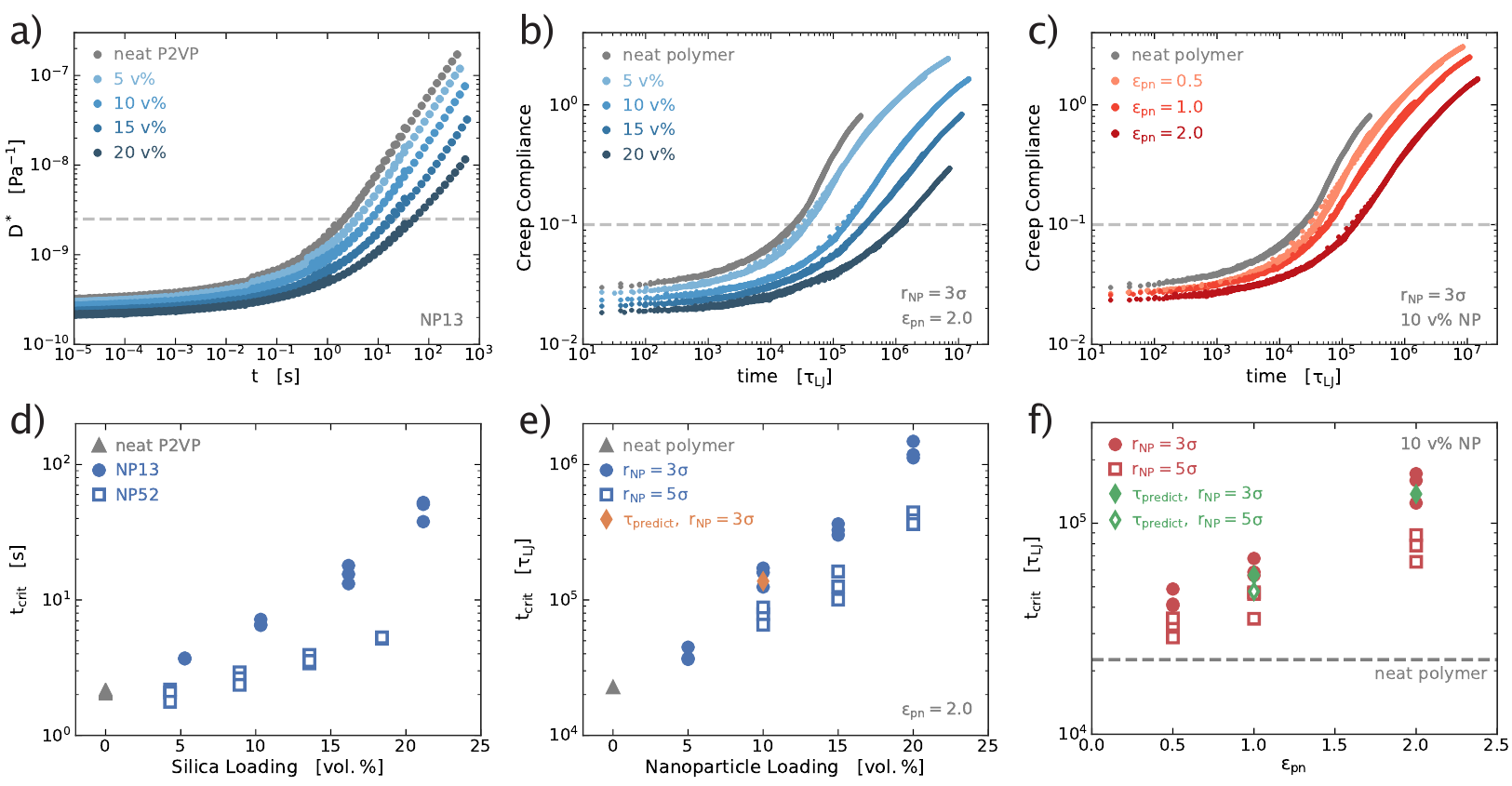}
    \caption{Characteristic creep compliance curves for several \textbf{a)} experimental and \textbf{b)} simulation composites containing NPs of strong polymer-NP interactions.
    The dashed line represents the critical creep compliance value.
    Critical deformation times as a function of NP loading for \textbf{d)} experimental and \textbf{e)} simulation composites.
    The qualitatively similar creep behavior in experimental and simulation systems suggests these simulations can accurately capture creep behavior in polymer nanocomposites.
    Results of different polymer-NP interactions in PNCs with $10\ vol\%$ NP are presented in \textbf{c)} and \textbf{f)} respectively.
    }
    \label{fig:comp_tcrit}
\end{figure*}

\textbf{Creep Attenuation} 
Representative master curves of the complex dynamic compliance ($D^*(t)$)) for the experimental and simulated PNCs are shown in Figures \ref{fig:comp_tcrit}a and \ref{fig:comp_tcrit}b for the smaller NP composites.
Master curves are shown with $T_{\mathit{ref}} = 105\ ^{\circ}C$ for experimental PNCs and $T_{\mathit{ref}} = 0.35$ for simulation systems.
At first glance, we observe several similarities in the compliance curves for the experimental and simulated PNCs.
In each case, the compliance in the glassy plateau modulus decreases with increasing NP loading and the fast creep regime (upturn in $D^*$) is shifted to longer times, demonstrating the reinforcement and creep attenuation abilities of NPs.

We define the critical deformation time as the time to reach a critical compliance value, $D^*_{\mathit{crit}}$, at $T_{\mathit{ref}}$.
For experimental systems, $D^*_{\mathit{crit}} = 2.5^{-9} Pa^{-1}$ (a 5\% strain under a $20 MPa$ stress) while for simulations $D^*_{\mathit{crit}} = 10^{-1}$ (a 4\% strain under a stress of $0.4$), as indicated by the dashed lines in Figure \ref{fig:comp_tcrit}a - \ref{fig:comp_tcrit}c.
Figures \ref{fig:comp_tcrit}d and \ref{fig:comp_tcrit}e plot $t_{\mathit{crit}}$ as a function of nanoparticle loading for all experimental and simulation systems.
In all cases, $t_{\mathit{crit}}$ increases exponentially with loading, with larger increases observed for smaller NP sizes, demonstrating how even modest loadings of small NPs with attractive polymer interactions can significantly suppress creep deformation.
This method for quantifying creep attenuation has been used previously and the exponential relation between critical time and NP loading are not qualitatively sensitive to the choice of critical compliance values.\cite{Buitrago2020} 
We note that while $t_{\mathit{crit}}$ is small ($< 100 s$) for this reference temperature, under ambient conditions (around {25 $^{\circ}C$}) {$t_{\mathit{crit}}$} may be on the order of several 10s to 100s of years due to the time-temperature equivalence in polymer viscoelasticity. 
These results are in qualitative agreement with previous measurements of creep in P2VP-silica nanocomposites.\cite{Buitrago2020}

We also study the effects of polymer-NP interactions on creep suppression in simulation systems, by fixing the NP loading at $10\ vol\%$ and varying the interaction parameters between polymers and NPs.
We find strongly interacting NPs better suppress creep, as evidenced by the shift of the upturn in compliance curves to a longer times and the increase in $t_{\mathit{crit}}$ with $\epsilon_{\mathit{pn}}$ (Figure \ref{fig:comp_tcrit}c and \ref{fig:comp_tcrit}f).

Overall, the qualitatively similar behavior observed in nanoparticle dispersion quality and creep attenuation for experimental and simulated strong-interaction PNCs suggests that the simulations performed in this work provide a good qualitative description of the behavior in the experimental composites and can be used to further probe the nanoscale behavior.

\textbf{Dynamical Decomposition in PNCs} 
In this work, we extend the dynamical decomposition developed in homogeneous glass-forming materials \cite{Schoenholz2016, Sussman2017} to PNCs, to reveal the mechanism of NP reinforcement under creep deformation.
A monomer's probability of rearranging at a given softness, $P_R(S)$, is used as the measurement of local polymer dynamics, and has been proven to be proportional to the inverse of the segmental relaxation time.\cite{Schoenholz2016,Schoenholz2017} 
Technical details of $P_R(S)$ calculation are presented in Methods.

In the work presented below, unless otherwise specified, we present the dynamical decomposition results for PNCs with $10\ vol\%$ NPs and neutral polymer-NP interactions for $T = 0.40$.
This analysis can be applied to other temperatures and all systems with dispersed NPs.
More details about other simulation conditions are provided in Table S1.
We begin with calculating the bulk-average dynamics in PNCs, $P_\mathit{R, avg}(S)$, by examining polymer monomers at least $8 \sigma$ away from one NP surface in the undeformed PNCs.

\begin{figure}[tb]
\centering
\includegraphics[width=1\linewidth]{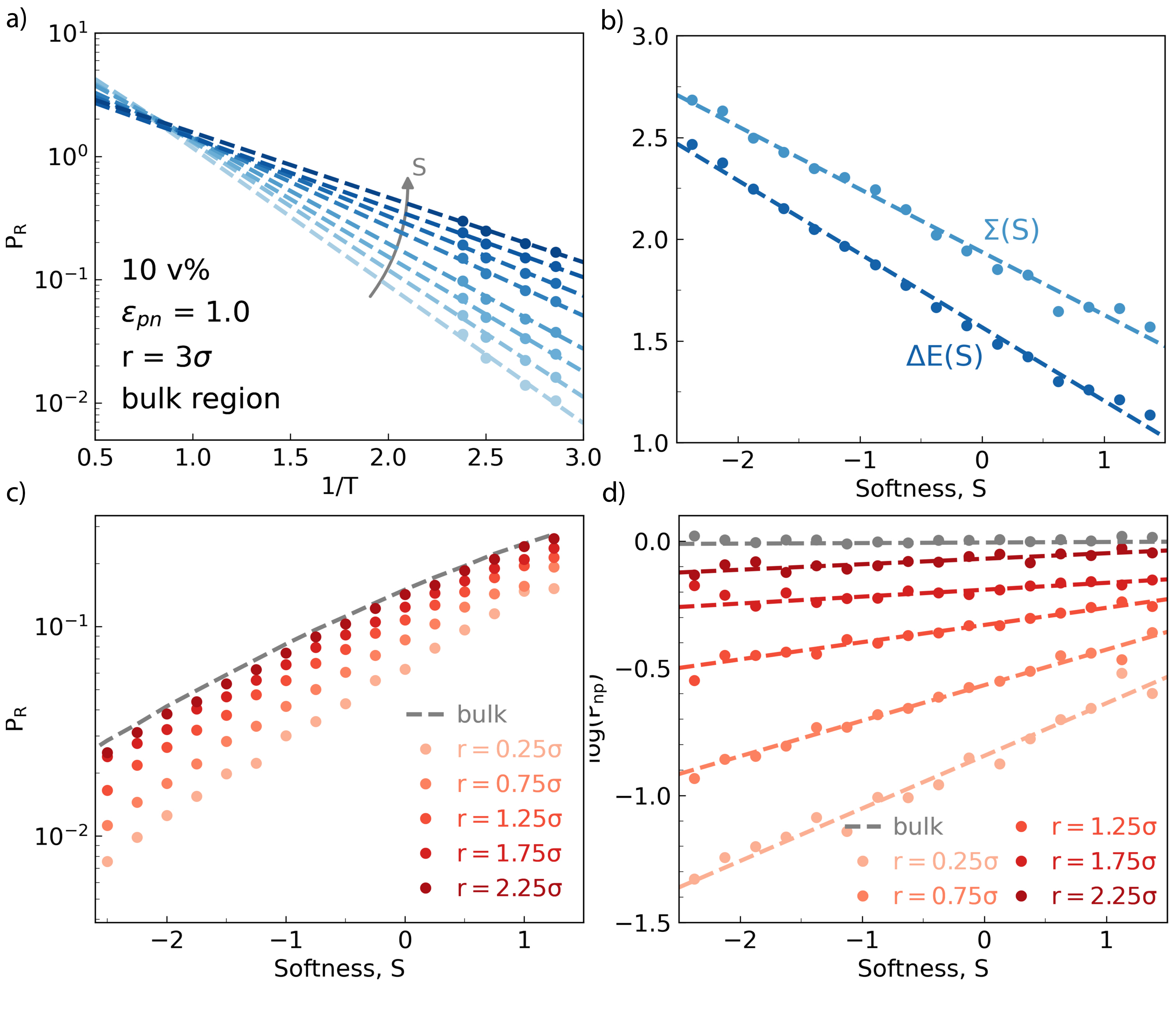}
\caption{Dynamical decomposition in PNCs.
\textbf{a)} Bulk-average probability of rearrangement for a given softness, $P_\mathit{R, avg}(S)$, as a function of $1/T$ at eight different softness values.
The color gradient represents the gradient in softness, ranging from light blue at $S = -2.75$ to dark blue at $S=1.25$.
\textbf{b)} The $\Delta E$ and $\Sigma$ values as a function of softness, calculated by fitting the results in a) to $P_\mathit{R, avg}(S) = \exp(\Sigma - \Delta E/T)$.
\textbf{c)} Isothermal probability of rearrangement as a function of $S$ at five distances from the NP surface ($r_\mathit{pos}$) and within the bulk-average region, at $T = 0.40$.
\textbf{d)} Isothermal $P_{\mathit{np}}$ as a function of $S$ at five $r_\mathit{pos}$ and the bulk-average region, for $T = 0.40$.
These simulations use $r = 3 \sigma$, $\epsilon_{\mathit{pn} = 1.0}$, and $10\ vol\%$ PNCs.}
\label{fig:PNC_dyn_decomp}
\end{figure}

In Figure \ref{fig:PNC_dyn_decomp}a, we plot $P_\mathit{R, avg}(S)$ as a function of $T^{-1}$ for 8 different softness values, ranging from $S = -2.75$ to $1.25$ (covering $97\%$ of polymer monomers).
For each softness, $P_\mathit{R, avg}(S)$ follows Arrhenius behavior and the left-extended fitting curves all intersect at the same point.
Furthermore, the softness-dependent activation energy exhibits a linear dependence on $S$ (Figure \ref{fig:PNC_dyn_decomp}b).
These two observations are similar to what has been seen in bulk glassy systems\cite{Schoenholz2016} and indicate that the bulk-average $P_R(S)$ in PNCs, can be expressed as the product of a structure-dependent and a structure-independent term,

\begin{equation}
\begin{aligned}
\label{eq:PR_bulk}
&P_\mathit{R, avg}(S) = \exp \left( \Sigma-\frac{\Delta E}{T} \right) = \exp \left( \Sigma_0-\frac{e_0}{T} \right)  \\
&\cdot \exp \left( -\left( \Sigma_1-\frac{e_1}{T} \right) S \right) = P_I(T) \cdot P_D(T,S),
\end{aligned}
\end{equation}

Here, $\Sigma$ and $\Delta E$ represent the entropic and enthalpic contribution to the energy barriers, respectively. $\Sigma_0$, $\Sigma_1$, $e_0$, $e_1$ are constants determined through linear fitting with softness, as shown in Figure \ref{fig:PNC_dyn_decomp}b, and are independent of temperature. Through the softness analysis, we can separate the structure-dependent components out from both $\Sigma$ and $\Delta E$ by combining terms involving $S$ (forming ($P_D(T, S)$)). While the other terms left form the structure-independent $P_I(T)$. 
The shared intersection point represents the temperature, $T_o$, where softness dependence of $\Sigma$ exactly cancels $\Delta E / T$ in Equation \ref{eq:PR_bulk}, which has been proven to scale with the onset temperature of glassy dynamics \cite{Schoenholz2016}. 

However, Equation \ref{eq:PR_bulk} breaks down near the NP interface because the probability of rearrangement at a given softness decreases as we approach the NP surface, as shown in Figure \ref{fig:PNC_dyn_decomp}c where $P_R(S)$ is plotted at five distances from the NP surface ($r_\mathit{pos}$) and within the bulk-average region, for $T=0.40$.
The downward shift in the $P_R- S$ curves leads to poor fit quality and no shared intersection point when plotting $P_R(S)$ as a function of $T^{-1}$ (see Figure S3).
The vanishment of shared intersection point further suggests the linear dependence between $\Sigma$ (and $\Delta E$) and softness is no longer valid near NP surface.
This is consistent with the intuitive expectation that $P_R$ should be lower for the \textit{same packing} (i.e., softness) when polymer monomers are near NPs, because NPs can slow nearby monomer dynamics.

To isolate the effect of NP proximity, we introduce a new quantity, $P_{\mathit{np}}$, defined as the ratio of $P_R$ over $P_{\mathit{R, avg}}$ for a given softness,
\begin{equation}
   \label{eq:P_np}
   P_{\mathit{np}}(r_{\mathit{pos}} | S_i) =\frac{P_R(r_{\mathit{pos}}, S_i)}{P_{\mathit{R, avg}}(S_i)}
\end{equation}
Thus, $P_{\mathit{np}}(S)$ is a measurement of NP slowing down effect on dynamics for a given structure.
We plot $P_{\mathit{np}}(S)$ at $T = 0.40$ for different distances from the NP surface in Figure \ref{fig:PNC_dyn_decomp}d and find that $P_{\mathit{np}}$ follows an exponential relation with $S$ at each distance, indicating that
$P_{\mathit{np}}$ can be expressed as $P_{\mathit{np}} = \exp[a_1(r_{\mathit{pos}}) \cdot S - a_0(r_{\mathit{pos}})]$, where both $a_0$ and $a_1$ depend on distance from the NP surface.
Note that, unlike $\Sigma_i$ and $e_i$, which are temperature independent, $a_i$ are temperature dependent (see Figures S2d--S2g).
This is consistent with the observation that thickness of the dynamically different region generally decreases with temperature.\cite{Zhang2019}
Recalling that $P_{\mathit{R, avg}}$ can be written as a product of $P_I(T)$ and $P_D(T,S)$ (Equation \ref{eq:PR_bulk}), this together gives the new expression for $P_R$ in PNCs near the NP surface:
\begin{small}
\begin{equation}
\begin{aligned}
\label{eq:P_R}
&P_R(r_{\mathit{pos}},S,T) = P_I^* \left(r_{\mathit{pos}},T\right) \cdot P_D^* \left(S, r_{\mathit{pos}}, T \right) = \exp \\
& \left(\Sigma_0 - \left( \frac{e_0}{T}+a_0 \right) \right) \cdot \exp 
\left( - \left(\Sigma_1 - \left(\frac{e_1}{T}+a_1 \right) \right) S \right)
\end{aligned}
\end{equation}
\end{small}
where $P_I^*$ and $P_D^*$ are the new expressions for the structure-independent and structure-dependent components, respectively.

Note that both $a_0$ and $a_1$ decays to zero in the bulk region, recovering to Equation \ref{eq:PR_bulk} and making this a general expression for polymer dynamics throughout PNCs. 
To our best knowledge, this is the first explicit expression between structure and dynamics on the monomer level in PNCs.

We want to emphasize that Equation \ref{eq:P_R} is not a combination of six random parameters aiming for better fitting, but with clear physical meanings embedded.
As shown in previous study\cite{Schoenholz2016}, $\Sigma_0$ and $e_0$ represent the structure-independent contributions in the entropic and enthalpic barriers for the monomer rearrangements. 
In contrast, $\Sigma_1$ and $e_1$ measure how sensitive these two barriers are to the local packing, respectively.
$a_0$ and $a_1$ are new parameters we introduce in this work. 
The former quantifies a constant slowing down for polymer dynamics carried by NPs, which only depends on distance to NP (at a given temperature), regardless of polymers' local structure. 
While the later $a_1$ counts the dependency of NP slowing down effect on the local packing.

\begin{figure}[tb]
    \centering
    \includegraphics[width=1\linewidth]{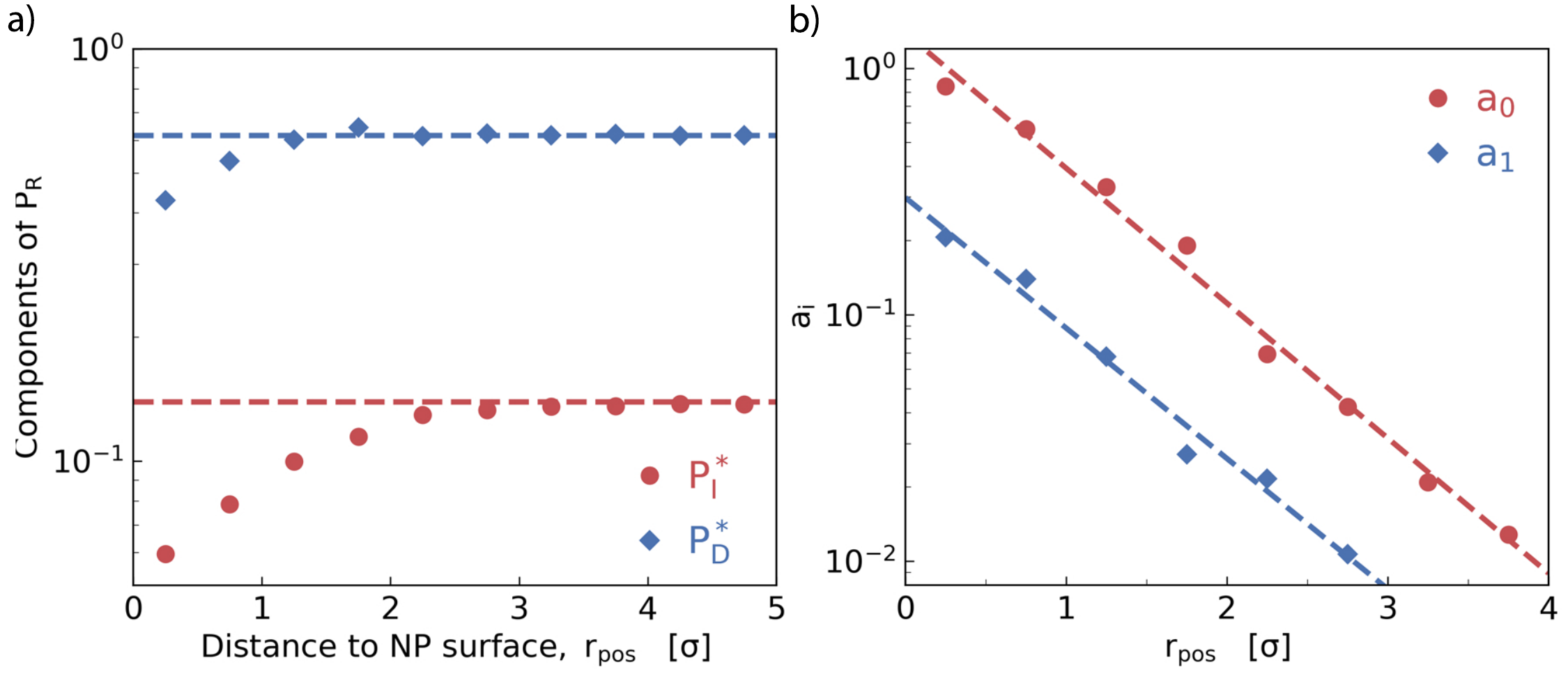}
    \caption{\textbf{a)} $P_I^*$ and $P_D^*$ as functions of the distance from the NP surface, $r_{pos}$, for $T=0.40$.
    The dash lines represent the bulk-average values of these quantities, $P_\mathit{I, avg}$ and $P_\mathit{D, avg}$.
    \textbf{b)} $a_0$ and $a_1$ as functions of $r_\mathit{pos}$, for $T=0.40$.
    The dashed lines are exponential fits.
    Both $a_0$ and $a_1$ decay exponentially with $r_\mathit{pos}$.
    These simulations use $r = 3 \sigma$, $\epsilon_{\mathit{pn}} = 1.0$, and $10\ vol\%$ PNCs.}
    \label{fig:PIPD_and_a0a1}
\end{figure}

In Figure \ref{fig:PIPD_and_a0a1}a, we plot $P_I^*$ and $P_D^*$ as functions of $r_\mathit{pos}$, both terms decrease near the NP surface, corresponding to slowed dynamics.
When calculating $P_D^*$, we use the average softness at that given $r_{pos}$.
The relative effect of the polymer-NP interface on $P_D^*$ and the thickness of the affected region are less than the effect of the interface on $P_I^*$, similar to the reported behavior near thin film interfaces.\cite{Sussman2017}
However, this behavior is not due solely to the presence of NPs, but also fluctuations in softness (packing) caused by polymer-NP interactions (see Figure S4a). The former effect can be quantified by $a_0$ and $a_1$, which both decay exponentially with $r_{pos}$ (see Figure \ref{fig:PIPD_and_a0a1}b). 
The exponential decay suggests that $\log(P_{np}) \propto \exp(-r_{\mathit{pos}})$, which also agrees with the 'double-exponential' relation of overall dynamical gradient near hard surface predicted in theory \cite{phan2020theory} and observed in simulations. \cite{scheidler2004relaxation, kob2012non, PhysRevE.89.052311, schweizer2019progress}

To determine if this decomposition could be expanded to PNCs under creep deformation, we deformed the $10\ vol\%$ NP composites using seven different stresses, $\sigma_c = 0.3$ to $0.9$, and measured the average softness as a function of time and $r_\mathit{pos}$ within both the constant strain rate regime and the full range of deformation (see Figures S5 and S6).
All $\sigma_c$ are lower than the composites' yield stress, which is greater than $1.0$ for all the systems.
Within the constant strain rate regime, the average softness was stable, following an brief jump caused by the initial elastic response.
A similar trend was reported in a recent work focused on softness analysis of colloidal gels under creep, where system average softness was found to depend on the strain rate.\cite{liu2021predicting}

Examining $P_R(S)$ as a function of $\sigma_c$ and $r_{\mathit{pos}}$ for both neutral and strong interactions reveals that the probability of rearrangement for a given softness varies significantly less with applied stress than with distance from the NP surface (see Figure S7).
Our recent work also prove that the stress-enhanced dynamics in polymer glasses can be described by an Eyring-like model, which buttresses the validity of dynamical decomposition under creep deformation. \cite{yang2021role}
These observations indicate that our new expression for polymer dynamics (Equation \ref{eq:P_R}) should still apply within the constant strain rate regime.
Thus, we perform the same dynamical decomposition analysis as carried out in Figure \ref{fig:PNC_dyn_decomp} for PNCs  under several different stresses ($\sigma_c = 0.3$ to $0.6$) and find Equation \ref{eq:P_R} remains valid (see Figures S8 and S9).

\begin{figure}[t!]
    \centering
    \includegraphics[width=1\linewidth]{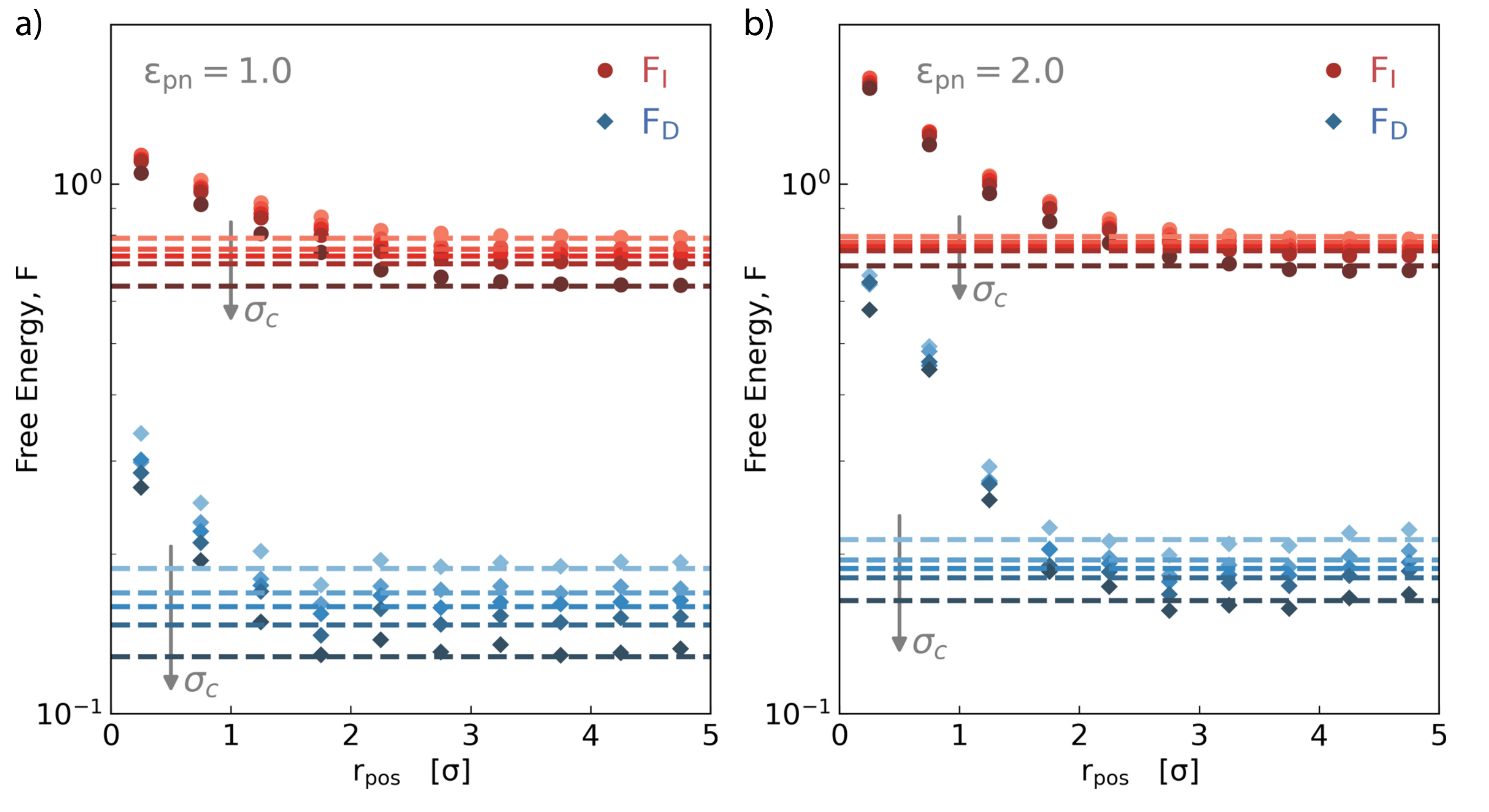}
    \caption{Structure-independent ($F_I$) and structure-dependent ($F_D$) free energy barriers as functions of $r_{\mathit{pos}}$ under different applied stresses for PNCs containing $10\ vol\%$ \textbf{a)} neutral polymer-NP interaction or \textbf{b)} strong polymer-NP interaction NPs.
    The color gradient represents the stress gradient, where the lightest color represents $\sigma_c=0$ and the darkest color represents $\sigma_c=0.6$. Dashed lines are the corresponding bulk-average values.
    These simulations use $r = 3 \sigma$ PNCs and $T=0.40$.
    }
    \label{fig:F_barrier}
\end{figure}

As suggested by previous work\cite{Schoenholz2016}, the exponential terms of $P_I^*$ and $P_D^*$ can represent the structure-dependent and structure-independent free energy barriers, $F_D$ and $F_I$ respectively, that need to be overcome for polymer monomers to rearrange. 
Thus, we can calculate the free energy barriers for monomer rearrangement during creep and the results of which are shown in Figure \ref{fig:F_barrier}.
Starting from the undeformed systems, free energy barriers near NPs are greater than the bulk-average energy barrier during deformation, and both $F_I$ and $F_D$ decrease with increasing stress.
Comparing the free energy barriers for PNCs with neutral and strong polymer-NP interactions, we observe a greater decrease in both $F_I$ and $F_D$ with stress for composites with neutral polymer-NP interactions, explaining the better creep suppression observed for PNCs with strong polymer-NP interactions.
The increased free energy barriers in PNCs with strong polymer-NP interactions hinder local polymer monomer rearrangements, thereby suppressing creep deformation.
We note that both $F_I$ and $F_D$ are just relative measurements of the barriers, whose magnitude can vary depending on the choice used to identify rearrangements (see Methods for more technical details). From our tests, the free energy barrier difference between $\sigma_c=0$ and $\sigma_c=0.6$ presented in Figure \ref{fig:F_barrier}a leads to an $47\%$ increase in total number of rearrangements.

On a high level, our results are also consistent with the recent advances of ECNLE theory, where Schweizer and coworkers show that glassy dynamics can be described by a combination of a local cage barrier and a long range elastic barrier\cite{Mirigian2013, Mirigian2015}. 
Both barriers are higher near rough surfaces \cite{phan2020theory} and decrease with external stress \cite{ghosh2020role}, while the long range elastic barrier is more sensitive to stress\cite{phan2019influence, phan2020theory, mei2021experimental}. 
However, the barriers in our model have a distinct microscopic origin. In ECNLE, the local cage and the elastic barrier are causally related, since the elastic barrier originates from the local cage expansion\cite{Mirigian2013}. 
While our approach using the machine-learned structure field, softness, enables us to effectively isolate the effect of structure, which should be a combination of the structural dependence in both the cage and elastic barriers proposed in ECNLE. 
In addition, our analysis automatically excludes the effect of surface-induced structure change, since all the analyses are carried on polymer monomers with same softness values.

\begin{figure}[t!]
    \centering
    \includegraphics[width=1\linewidth]{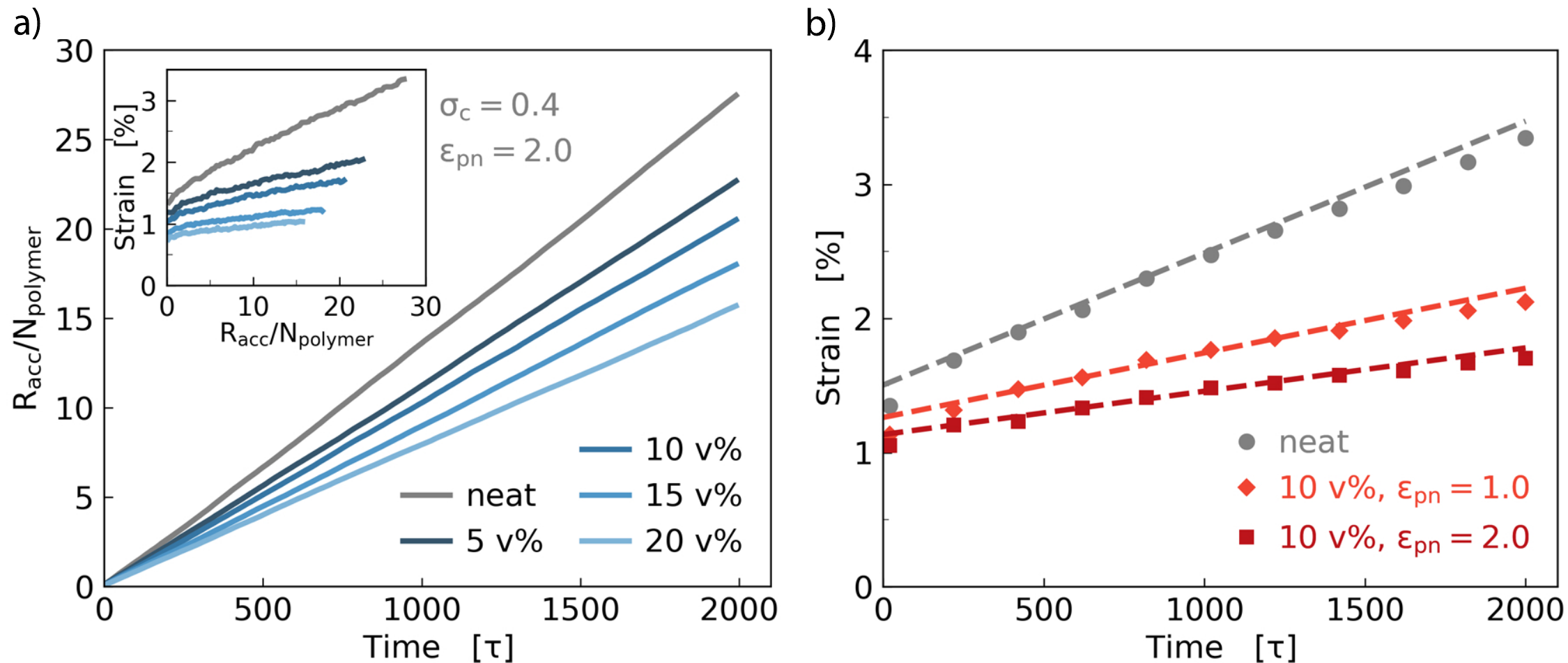}
    \caption{\textbf{a)} Normalized $R_{\mathit{acc}}$, as a function of time, for different NP loadings at $T = 0.40$ and $\sigma_c = 0.4$.
    The inset shows the corresponding strain versus $R_{\mathit{acc}}/N_{\mathit{polymer}}$ curves.
    \textbf{b)} Predicted strain-time responses (dashed lines) of three systems at $T = 0.40$; dots are the measured strain.
    These simulations use neat polymers or $r = 3 \sigma$ PNCs.
    }
    \label{fig:R_acc_strain}
\end{figure}

\textbf{Predicting Strain Response}
Plastic deformation in disordered solids happens through local structural rearrangements. \cite{PhysRevE1998}
Thus, to connect the macroscopic mechanical response to microscopic polymer dynamics, we can examine the relationship between strain, $\epsilon$, and the accumulated number of monomer rearrangements, $R_{\mathit{acc}}$, within the constant strain rate regime.
In Figure \ref{fig:R_acc_strain}a, we plot $\overline{R_{\mathit{acc}}}$ ($R_{\mathit{acc}}$ normalized by the total number of polymer monomers, $N_{\mathit{polymer}}$) with time and find they follow a intriguing linear relationship, with the slope decreasing as NP loading increases.
In other words, the ratio of monomer rearrangements to $N_{\mathit{polymer}}$ remains almost constant during deformation and is determined by the NP loading (see Figure S10).
As we are within the constant-strain-rate regime, this implies the strain, $\varepsilon$, also increases approximately linearly with normalized $R_{\mathit{acc}}$ after a brief initial transient, as shown in the inset of Figure \ref{fig:R_acc_strain}a. Therefore, $\varepsilon$ can be estimated from $\overline{R_{\mathit{acc}}}$ using:
\begin{equation}
    \label{eq:strain}
    \varepsilon = k_1 \overline{R_{\mathit{acc}}} + k_0
\end{equation}
Here, $k_0$ and $k_1$ are constants obtained through linear fitting, representing initial elastic response and rearrangements needed to reach $1\%$ strain respectively. This linear relationship suggests that we can predict strain as a function of time from the structure information (i.e. softness), because $R_{\mathit{acc}}$ can be estimated by integrating $P_R$,
\begin{equation}
    \label{eq:Racc_brief}
    \overline{R_{\mathit{acc}}} = \frac{1}{N_{\mathit{polymer}}} \int_{0}^{time} \int_{0}^{\infty} \int_{S_0}^{S_\infty} N_s P_R\  dS dr_{\mathit{pos}} dt.
\end{equation}
Here, $N_s$ is the number of monomers with a given softness and $r_\mathit{pos}$.
$P_R$ is the probability of rearranging and can be calculated through our dynamical decomposition model (Equation \ref{eq:P_R}) given the system's softness distribution.

Having connected the number of rearrangements required to reach a given strain, we can move to predict the strain response from $P_R(S)$. We find that the softness distributions remain unchanged after the initial elastic response and the magnitude of the softness change due to the elastic response grows linearly with stress.
Therefore, the softness distribution during creep can be estimated from the pre-deformation sample, enabling us to directly predict strain responses within the low strain regime where the strain rate is approximately constant and spatially homogeneous from the structure in the undeformed system (see more technical details in Figures S11 and S12).

In Figure \ref{fig:R_acc_strain}b, we plot the measured strain (dots) and the predicted strain (dashed lines) as functions of time, for three systems at $T = 0.40$, finding excellent agreement.
Strain predictions for other temperatures are also accurate and can be found in Figure S13.
These predicted strains, combined with the shift factors from TTS, can be used to estimate the critical deformation time of different composites.
In Figures \ref{fig:comp_tcrit}e and \ref{fig:comp_tcrit}f we predict $t_\mathit{crit}$ for multiple composite systems (diamond points) and demonstrate excellent agreement with the measured values.
This strain prediction provides a microscopic picture of the glassy dynamics that leads to creep. 
By connecting the monomer-level structure to molecular rearrangements, the model presented above that decomposes the dynamics into a product of a structure-dependent and a structure-independent process that can be used to predict the creep response. This approach can be applied not just in neat polymers but also polymer nanocomposites. Future work will seek to understand how the picture changes when the relationships between stress, strain, and softness become nonlinear. 

\section{Conclusions}

In this work, we studied the ability of well-dispersed NPs to suppress creep using both simulations and experiments, demonstrating qualitative agreement. 
We find that a composite's critical deformation time increases exponentially with NP volume fraction in both simulation and experiments, with smaller NPs having a larger effect. 
Simulation results also suggest that this exponential dependence remains for different polymer-NP interactions, while stronger interactions can better suppress creep.

We speculate that this exponential dependence is a collective outcome of several factors including the exponential decay of NP slowing down effect on dynamics versus $r_{\mathit{pos}}$ (described by $a_1$ and $a_0$), the increase of interfacial polymer ratio versus NP loading, the magnitude of the change in the TTS shift factors with changing NP loading, and the shift in softness distribution caused by both NPs and stress. 
The first three factors help suppress creep, while the latter one depends on specific situation. 
Because NPs usually reduce softness but stress tends to increase it, and a decrease in softness would promote the suppression. 
These factors are all included in our later prediction of strain response from static structure information. 
The agreement between strain prediction and direct measurements in simulation further support the change of critical time cannot be attributed to a simple factor. 
Our primary tests suggest that the relative contribution of different factors depends on system conditions (polymer-NP interactions, NP loadings, etc.) and a systematic study by isolating these factors respectively is needed for the future work. It would also be insightful to experimentally study PNCs with significantly different softness, either near nanoparticle surfaces or for the polymer matrix itself.

Our dynamical decomposition model provides an explicit relationship between structure and dynamics on the particle level in PNCs, especially near the NP surface.
Our results suggest that, in addition to the structure change, the modified structure-dynamics relation ($P_R(S)$) is also responsible for the slow interfacial dynamics. 
The change of $P_R(S)$ can be described by the product of two processes ($P_I$ and $P_D$), one is dependent on polymer packing (softness) and the other is not.
Both processes depend on the distance to the NP surface, $r_{\mathit{pos}}$, while the former one increases with softness exponentially at each $r_{\mathit{pos}}$.
For the same polymer packing, the NP slowing down effects on dynamics ($a_i$) decay exponentially with $r_{\mathit{pos}}$ in both components ($P_I$ and $P_D$). These together constitute the 'double exponential' dynamical gradient predicted and observed before \cite{scheidler2004relaxation, kob2012non, PhysRevE.89.052311, schweizer2019progress, phan2020theory}.
The existence of the structure-independent process also explains why the thickness of the dynamically-distinct region around NPs can differ from the thickness of the region with an altered monomer structure, as reported in previous studies\cite{Hanakata2015, Zhang2019}.

Our free energy barrier analysis demonstrates that NPs suppress creep in PNCs by increasing both the structure-dependent ($F_{D}$) and structure-independent ($F_{I}$) free energy barriers for polymer motion.
The former is described by $a_1$ and shifts in the softness distribution (changes in particle packing), while the later is captured by $a_0$.
Thus, the increase in polymer packing density near NPs is not the only source of NP reinforcement. 
In contrast, the presence of NPs alters the relationship between structure and glassy dynamics ($a_1$) and also provides additional slowing down effect through a structure-independent process ($a_0$).
It is tempting to speculate that $a_1$ and $a_0$ are due to the presence of a stiffer phase (rigid NPs), which increases the local stiffness of the interfacial layer and blocks potential rearrangements in certain directions.
In addition, when comparing the role of polymer-NP interactions, we find that the better creep suppression of the strong NPs is due to the higher energy barriers near NP surface and the reduced sensitivity of these barriers to increasing stress.

The dynamical decomposition model also enables us to predict the overall strain response within the low strain regime and, thus, the constant-strain-rate response limit of a given PNC directly from the structure information of the pre-deformation sample. We believe it provides additional insights in the mechanism of creep suppression and will potentially reduce development time when designing and screening PNCs for structural applications. Further, the connection built between particle-level softness and the overall mechanical response suggests that softness plays an important role as a structural descriptor for the development of constitutive models of glasses that broadly predict non-equilibrium behavior.\cite{ottinger2005beyond}

\section{Methods and Materials}
\textbf{Simulation methods}
All simulations were performed using the LAMMPS molecular dynamics package.\cite{Plimpton1995}
A coarse-grained bead-spring model was used to construct the polymer matrix.\cite{Kremer1990}
Each simulation system contains 405 monodispersed polymer chains consisting of 128 Lennard-Jones (LJ) interaction sites, connected by flexible harmonic bonds for a total of 51,840 polymer particles.
The standard 12-6 Lennard-Jones cut potential is used to describe all non-bonded monomer interactions,
\begin{equation}
\begin{aligned}
    \label{eq:LJ}
    U^{nb}(r_{ij}) = 4\epsilon_{ij} &\left[ \left( \frac{\sigma}{r_{ij}} \right)^{12} - \left( \frac{\sigma}{r_{ij}} \right)^{6} \right] - U_{\mathit{cut}}, \\
    &\qquad r_{ij} < 2.5 \sigma
\end{aligned}
\end{equation}
where $U_{\mathit{cut}}$ is the value of the 12-6 potential at our cut-off distance, $r_c = 2.5 \sigma$, and $\sigma$ is the bead size.
Both polymer-polymer ($\epsilon_{pp}$), and NP-NP ($\epsilon_{nn}$) interactions are fixed at 1.0, while polymer-NP ($\epsilon_{pn}$) interactions are set at 0.5, 1.0, and 2.0 representing weak, neutral, and strong interactions respectively.
Bonded monomer interactions are described by the harmonic bonding potential,
\begin{equation}
    \label{eq:harmonic}
    U_{ij}^b = K \left( r - \sigma \right)^2
\end{equation}
where $K=400\epsilon/\sigma^2$, and $\sigma$ is the diameter of the monomers.
All units for quantities taken from the simulation are in LJ-reduced unit notation.
The reduced temperature, $T$, is expressed as $T = kT^* / \epsilon$, and the LJ-time, $\tau_{\mathit{LJ}} = t^* \sqrt{\epsilon / m \sigma^2}$, where $k$ is the Boltzmann constant, $m$ is the mass of a single LJ interaction site, $T^*$ is temperature, and $t^*$ is time.
The asterisk indicates quantities in laboratory units.

The NPs are modeled as amorphous, solid, and rigid spheres of LJ sites, cut from a bulk LJ liquid at high temperature ($T = 10.0$) and high density ($\rho_0 = 1.25$) with nominal radii, $R_p$, of $3.0$ and $5.0 \sigma$ for small and large NPs, respectively.
The resulting NP has the same density as the high temperature, high pressure LJ liquid, ensuring an amorphous NP.
Due to its amorphous nature, the actual radius can be slightly smaller than the nominal radius in some parts of the NP.
For simplicity, we used the nominal radius to define the position of the NP surface.

In addition to the neat polymer systems, we prepared PNCs with four different NP volume fractions, $5v\%$, $10v\%$, $15v\%$, and $20v\%$.
Strong polymer-NP interactions were simulated at all NP loadings, while additional weak and neutral interactions were simulated at $10v\%$.
More details about simulated composites can be found in Table S1.

Systems were equilibrated in the NPT ensemble at $T=1.0$ and $P=0$ with a timestep of $0.002 \tau_{\mathit{LJ}}^{-1}$.
Connectivity altering Monte Carlo moves (i.e. bond swap) were applied for the purpose of reaching equilibrium.\cite{karayiannis2002novel,banaszak2003new,auhl2003equilibration}
For a given system, different configurations are separated by at least one polymer diffusion time, $\tau_D$, to guarantee their independence.
Next, all the PNCs were quenched to different target temperatures ($T = 0.2$ to $0.5$) with a cooling rate of $\Gamma = 10^{-4} \tau_{LJ}^{-1}$, followed by an aging period of $10^4 \tau_{\mathit{LJ}}$.
Creep deformations at a series of stresses ($\sigma_c = 0.3$ to $0.9$) were then performed at different temperatures ($T = 0.35$ to $0.50$).
For each configuration, the system was uniaxially deformed in each dimension while maintaining constant pressure in the transverse directions and the strain responses were averaged.
Compliance vs time curves at each temperature were shifted using time temperature superposition (TTS) to generate a master curve at $T_{\mathit{ref}} = 0.35$.

\textbf{Softness calculation}
The softness of a polymer monomer refers to the probability that the monomer will rearrange over a given time period.
To calculate softness, we employ a set of structure functions and form a  feature vector to represent a polymer monomers' local environment.
In other words, the structure functions describe the distribution of other monomers around a central monomer.
These feature vectors represent the monomers' coordinates in high dimensional space (with the dimension equal to the number of structure functions).
By applying the Support Vector Machine (SVM) machine learning algorithm, we can find a hyperplane that best separates the rearranging monomers from the non-rearranging monomers.
We then define softness as the particle's signed distance to this hyperplane, with a positive value corresponding to the rearranging side of the hyperplane and a negative value to the non-rearranging side.
Here we used the hyperplane trained in our previous work,\cite{Yang2020} which was trained on a quiescent neat polymer system with two groups of structure functions for each polymer monomer i. The first group,
\begin{small}
\begin{equation}
\begin{aligned}
    \label{eq:G_R}
    G_R(i; \mu,L) &= \sum_j \max \ [ \exp \ ( -\ ( R_{ij}-\mu \ )^2 / L^2 ) \\
    & - \epsilon_R, \ 0],
\end{aligned}
\end{equation}
\end{small}
\\describes the radial structural characteristics while the second
\begin{small}
\begin{equation}
\begin{aligned}
    \label{eq:G_A}
    &G_A(i;\xi,\lambda,\zeta) = \sum_j \sum_k \max\  [ \exp (-( R_{ij}^2 + R_{ik}^2 \\
    & + R_{jk}^2 ) \ / \ \xi^2 ) \cdot \ ( ( 1 + \lambda \cos \theta_{ijk} \ ) / 2 )^\zeta - \epsilon_A, 0 ]
\end{aligned}
\end{equation}
\end{small}
\\describes three-body orientation characteristics.

Here, $R_{ij}$ is the distance between particle i and particle j; $\theta_{ijk}$ is the angle between particle $i$, $j$, and $k$; and $\mu$, $L$, $\xi$, $\lambda$, $\zeta$ are all parameters varied to construct different structure functions.

Given the two types of particles in our systems, polymer and NP sites, we make an approximation by treating the NP sites the same as the polymer sites; this avoids creating a region of artificially high softness near the NPs due to a decrease in monomer density.
Our recent work shows this method is able to probe the structure change near the NP surface.\cite{Yang2020} In the supporting materials, we provide a detailed discussion and justification of this assumption.

\textbf{Dynamical measurement}
For our dynamical decomposition model, we use $P_R(S)$, the probability of rearrangement for particles with a given softness, as the measurement of polymer dynamics. 
This method has been proven to be efficient and can predict particles' relaxation time robustly.\cite{Schoenholz2016,Schoenholz2017} 
Details of $P_R(S)$ calculation are presented below.

Since we are interested in composites under creep deformation, we use the quantity $D^2_{\mathit{min}}$ as recommended in literature to determine whether a monomer is rearranging,\cite{PhysRevE1998} accounting for the monomer's non-affine motion. $D^2_{\mathit{min}}$ is defined as
\begin{equation}
    \label{eq:D2_min}
    D^2_{\mathit{min}}(i;t) = \frac{1}{N_i} \sum_j^{N_i} \left[ \vec{r}_{ij} (t+\delta t) - \vec{\Lambda}_{i}(t) \vec{r}_{ij}(t) \right]^2,
\end{equation}
where $\vec{r}_{ij}$ is the displacement vector between particle $i$ and $j$ at time $t$ and $\vec{\Lambda}_{i}(t)$ is the best fit local gradient tensor for particle $i$ that minimizes $D^2_{\mathit{min}}(i;t)$.
A monomer is considered to be rearranging if its value is larger than $D^2_\mathit{min, 0} = 0.1$.
As shown in both previous work and our own tests,\cite{Sussman2017,Yang2020} we find our results are qualitatively insensitive to the choice of $D^2_\mathit{min, 0}$ over a reasonable range of values ($0.06$ to $0.23$).

After identifying the rearranging particles, we then bin both rearranging particles and all particles based on their softness. $P_R(S)$ can be calculated by dividing the number of rearranging particles by the total number of particles with that softness.

In this work, we used softness values ranging from $-2.75$ to $1.25$ (covering $97 \%$ of monomers in the system) with a resolution of 0.25 (our softness distribution follows a normal distribution with $\sigma_S = 0.92$) when calculating $P_R(S)$.

\textbf{Materials}
Commercial grade poly(2-vinylpyrridine) (P2VP, $M_n = 70\ kg\  mol^{-1}$ and $M_w / M_n = 2.4$) was obtained from Scientific Polymer Products, Inc. (Ontario, NY).
Silica nanoparticles dispersed in 2-butanone were obtained from Nissan Chemical (MEK-ST and MEK-ST-L).
The nanoparticles diameters were measured via small angle X-ray scattering (SAXS) and were found to be log-normally distributed with average diameters of $13 nm$ and $52 nm$, respectively, and polydispersities of $0.30$ and $0.29$.
For clarity, the small nanoparticles will be referred to as NP13 and the large nanoparticles as NP52.
Nanoparticles were transferred from 2-butanone to methanol via a solvent exchange process using hexane to crash the nanoparticles out of solution.

\textbf{Polymer Nanocomposite Preparation}
The P2VP was dissolved in methanol at a concentration of approximately $50\ g/L$.
Pentaerythritol tetrakis (3-(3,5-di-tert-butyl-4-hydroxyphenyl) propionate) (Sigma-Aldrich), an antioxidant, was added to the polymer solutions at a concentration of $0.1\ w\%$ of the polymer mass to prevent polymer decomposition during nanocomposite preparation and thermal processing.

The nanoparticle solutions were diluted to approximately $15\ g/L$ with methanol prior to adding, dropwise, to the P2VP solutions to target nanoparticle loadings equal to $5\ vol\%$, $10\ vol\%$, $15\ vol\%$, and $20\ vol\%$.
The nanocomposite solutions were mixed for $24 \ h$ before casting on PTFE dishes at $120\ ^{\circ}C$ to remove the majority of the solvent.
The samples were then transferred to a vacuum oven where they were annealed for $24 \ h$ at $150\ ^{\circ}C$ to remove the remaining solvent.

After thermal annealing, the composites were hot pressed in aluminum molds at $150\ ^{\circ}C$ under $0.5$ metric tons for $10$ mins, creating samples with nominal dimensions of $35 \times 3.0 \times 0.5 \ mm$.
Samples were quenched and removed from the molds before annealing for $12 \ h$ at $120\ ^{\circ}C$ under vacuum followed by a slow cool in the vacuum oven to a temperature of less than $40\ ^{\circ}C$.
This process was used to remove residual stresses caused by hot pressing and ensure a standard thermal treatment for each set of samples.
The hot pressed samples were used for all subsequent analysis.

\textbf{Thermogravimetric Analysis (TGA)}
Nanoparticle loadings were determined via TGA using a TA Instrument Q600 SDT.
For each sample, $~10\ mg$ of composite was heated to $150\ ^{\circ}C$ at a rate of $10\ ^{\circ}C$, held for $20$ mins, and then heated to $900\ ^{\circ}C$ at a rate of $20\ ^{\circ}C \ min^{-1}$ under flowing air.
The volume percent of silica was calculated using a silica nanoparticle density of $2.2\ g \ cm^{-3}$ and a P2VP density of $1.2\ g \ cm^{-3}$.
A summary of the composite loadings is provided in Table S2.

\textbf{Temperature Modulated Differential Scanning Calorimetry (TMDSC)}
The glass transition temperature ($T_{\mathit{g}}$) of the nanocomposites was measured via TMDSC using a TA Instrument Q2000. Measurements were made upon cooling a $~10\ mg$ sample at a rate of $5\ ^{\circ}C \ min^{-1}$ with a modulation time of $30 \ s$ and an amplitude of $\pm 0.5\ ^{\circ}C$ from $40\ ^{\circ}C$ to $170\ ^{\circ}C$.
$T_{\mathit{g}}$ was defined as the inflection point of the heat flow thermograms.
$T_{\mathit{g}}$ of the composite systems was within $1\ ^{\circ}C$ of the neat polymer $T_{\mathit{g}}$. See Table S2 for a full list of $T_{\mathit{g}}$.

\textbf{Small Angle X-ray Scattering (SAXS)}
Particle dispersion was examined using SAXS performed on a Xenocs Xeuss 2.0 with a GeniX3D copper source ($8 \ keV$, $1.54$ \AA) and a PILATUS3 1M detector.
Sample to detector distances of $1.2\ m$ and $6.4\ m$ were used, corresponding to a wave vector (q) range of around $0.002$ \AA$^{-1}$ to $0.2$ \AA$^{-1}$.
Two-dimensional scattering patterns were azimuthally integrated to one dimension using the Foxtrot software.
Nanoparticle form factor scattering was measured using a dilute solution of nanoparticles in methanol ($0.1\ vol\%$ and $0.5\ vol\%$ for NP13 and NP52 nanoparticles, respectively) within a low-noise flow cell (Xenocs).
The form factors were fit with a polydisperse sphere model using the SASView analysis software to determine average particle size and polydispersity.

\textbf{Scanning Electron Microscopy (SEM) \& Focused Ion Beam (FIB)}
PNC samples were fractured at room temperature to expose a fresh interior region.
Cross-sectioning for two-dimensional analysis was performed using a Zeiss Crossbeam 540 FIB/SEM.

\textbf{Dynamic Mechanical Analysis (DMA)}
Oscillatory DMA measurements were performed on a TA Instruments RSAIII in tension mode.
Strain-controlled frequency sweeps ($0.03$ Hz to $30$ Hz) were performed with a $0.05\ \%$ strain (within the linear viscoelastic regime) in $5\ ^{\circ}C$ increments between $30\ ^{\circ}C$ and $110\ ^{\circ}C$.
Three samples were measured for each composite system and the results were averaged.
DMA measurements were shifted using time-temperature superposition with a reference temperature of $105\ ^{\circ}C$.
The dynamic modulus, $E^*(\omega)$, obtained from the oscillatory DMA experiments (where $\omega$ is the frequency of the dynamic loading), was transformed to transient compliance, $D^*(t)$, using known analytical relationships.\cite{Honerkamp1993a,Riande1975}
As a first approximation, we used $D^* = 1 / E^*$ and $t \approx 1 / \omega$.
This method has been used previously to examine creep in polymer nanocomposites.\cite{Buitrago2020}

\section{Supporting Information}
Supporting information includes: \\
Figures S1 - S15 \\
Tables S1 - S3 \\
Supplementary text \\
SI Reference

\begin{acknowledgement}

The authors thank support from ExxonMobil Research and Engineering, in addition to DOE-BES via DE-SC0016421. The authors acknowledge use of the Dual Source and Environmental X-ray Scattering facility operated by the Laboratory for Research on the Structure of Matter at the University of Pennsylvania (NSF MRSEC 17-20530).
The equipment purchase was made possible by a NSF MRI grant (17-25969), a ARO DURIP grant (W911NF-17-1-0282), and the University of Pennsylvania. The authors also acknowledge computational resources provided by Extreme Science and Engineering Discovery Environment (XSEDE) via allocation TG-DMR150034, which is supported by National Science Foundation grant number ACI-1548562.

\end{acknowledgement}




\bibliography{achemso-demo}

\end{document}